# Probing anisotropy in a new noncentrosymmetric superconductor BiPd

Bhanu Joshi*, A. Thamizhavel, and S. Ramakrishnan

*Department of Condensed Matter Physics and Materials Science,
Tata Institute of Fundamental Research, Colaba, Mumbai – 400 005, India*

*E-mail: bpjphy@gmail.com*

Strength of Rashba type antisymmetric spin orbit coupling has the crucial effect on superconducting and normal state properties in noncentrosymmetric (NCS) superconductors. This can be seen on various anisotropic properties of the system. For a system with lowered crystalline symmetry as in noncentrosymmetric crystals, the superconducting pairing state can have correlation to the normal state properties. In our earlier work we have shown that α-BiPd which has a monoclinic ($P\_2_1$) structure with the absence of the centre of inversion exhibits bulk superconductivity at 3.8 K. In order to understand the role of spin-orbit scattering, we have studied the anisotropic properties of high quality single crystal (RRR~170) of NCS BiPd by measuring magnetic susceptibility, electrical transport from 1.8 K to 300 K and isothermal M_H loop (1.8 K to 3.8 K) in the superconducting state for <010> and <100> crystallographic directions respectively. Moderate anisotropy up to 30 % has been found in the electrical transport possibly indicating significant Rashba type antisymmetric spin orbit coupling present in α-BiPd.

## 1. Introduction

Noncentrosymmetric superconductors (NCS) are extremely interesting systems in current research as they can have rare coexistence of spin-singlet and spin-triplet superconducting pairing states. For noncentrosymmetric system there exists Rashba type antisymmetric spin-orbit coupling (ASOC) due to the lack of inversion symmetry and depending upon its strength, it can affect superconducting as well as normal state properties of the system [1, 2, 3]. Strength of ASOC is directly proportional to the square of effective atomic number ($Z^2$) and electronic wave vector *k* as seen by electrons within a particular band, so one can always expect high strength of Rashba type ASOC for materials containing high Z elements. With sufficient strength ASOC can induce mixing of singlet and triplet pair components in superconducting state and Fermi liquid instability in normal state. Recently some groups have also envisaged the manifestation of protected edge states within gapless topological phases and further realization of elusive Majorana fermions at SC vortex core in NCS [4, 5, 6]. To really observe such titillating effects of missing inversion symmetry one need a high quality single crystal so that sensitive detrimental effect of disorder towards triplet pair component should be minimized. It would be ideal if in addition one could choose a system where additional complications of strongly correlated electrons on superconducting properties can be avoided.

Monoclinic BiPd is one such sample which surprisingly fulfills all above mentioned tough prerequisites in totality to unravel the hidden enriched physics of NCS. BiPd turns out to be a weakly correlated electron system (non-heavy fermion) [7]. The high quality single crystalline nature (RRR~170) [7] makes its superconducting state unaffected by disorder. This is the main reason that recently some of us have already observed possibility of triplet pairing in BiPd [8, 9]. This signature of triplet pairing in BiPd can be ascertained more profoundly by coherently converging results of other independent microscopic measurements. They are in progress.

Strength of the ASOC can be reflected on various anisotropic properties of the system. In this present work we have measured the anisotropy via transport and magnetization measurements for different crystallographic directions of BiPd NCS single crystal.

## 2. Experiments

*2.1 Anisotropy in M_H loop for BiPd in the superconducting state:*

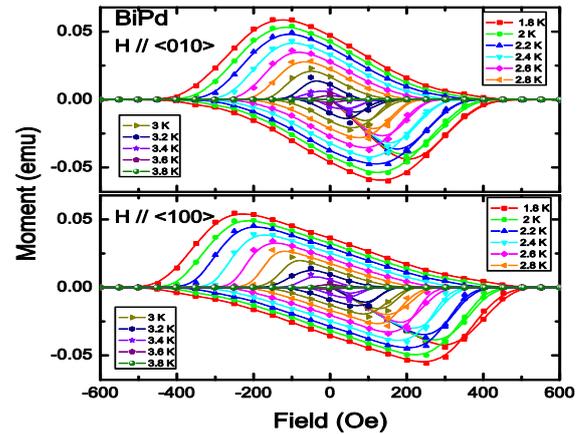

Isothermal Magnetization vs Field loops for a rectangular bar (mass 23.1 mg, size 3.8 mm × 1.25 mm × 0.5 mm) of BiPd has been recorded from 1.8 to 3.8 K at every 0.2 K interval with magnetic field along <010> and <100> crystallographic directions, respectively. Significant anisotropy in the upper critical field ($H_{c2}$) values has been observed. At 1.8 K, $H_{c2}$ is found to be 920 Oe for H // <010> and 1380 Oe for H//<100> direction respectively, revealing nearly 33 % anisotropy in the upper critical field for BiPd at this temperature. Such directional anisotropy in upper critical field is frequently found for many of the unconventional high Tc superconductors.

**Fig. 1.** Isothermal M_H loop (1.8 K to 3.8 K) for BiPd in the superconducting state with magnetic field along <010> and <100> crystallographic directions respectively.

*2.2 Anisotropy in the normal state susceptibility:*

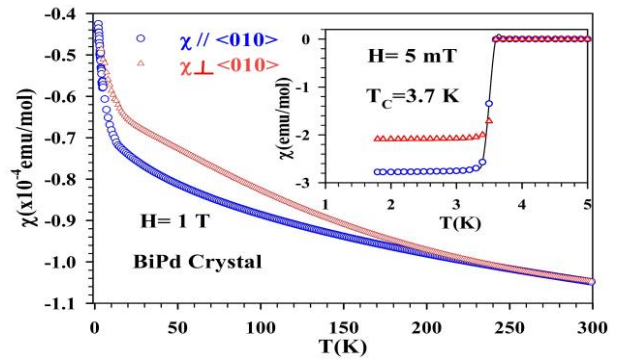

Normal state magnetic susceptibility of BiPd has been measured under the magnetic field of 1 tesla (sufficiently above the upper critical field of BiPd) for magnetic field parallel and orthogonal to <010> direction. Since the experimentally observed susceptibility is combined manifestation of Pauli susceptibility, Landau susceptibility and core electron susceptibility. Presence of weak room temperature diamagnetism of BiPd can be ascribed to the major core electron contribution from Bi as explained in Ref [7].

**Fig. 2.** Temperature dependence of normal state magnetic susceptibility (at 1 Tesla magnetic field) of BiPd from 1.8 K to 300 K for magnetic field parallel and orthogonal to <010> direction respectively. Inset is showing clear superconducting transition below 3.7 K at 5 mT field for both the directions.

$$\chi_{Observed} = \chi_{Landau} + \chi_{Pauli} + \chi_{Core} \,. \qquad (1)$$

Anisotropy in the normal state susceptibility which starts below 200 K and gets reduced below 5 K is not understood now as it cannot be explained in terms of isotropic core electrons susceptibility and it possibly reflects the anisotropic temperature dependence of the density of states at the Fermi level itself.

*2.3 Anisotropy in transport:*

Resistivity measurement has been done for current parallel and orthogonal to <010> direction. The magnitude of anisotropy in BiPd keeps on increasing with increasing temperature and moderate anisotropy up to 30 % has been observed at room temperature suggesting the presence of significant spin-orbit coupling strength present within BiPd.

Low temperature part (5 K to 30 K) of normal state resistivity (Fit 2) has been fitted well by the expression
$$\rho(T) = \rho_0 + A\,T^n. \qquad (2)$$
for both the directions with temperature exponent n close to 3. This $T^3$ dependence of resistivity is less common in transition metals and explained by Wilson's model [10] which considers interband s-d phonon induced scattering.

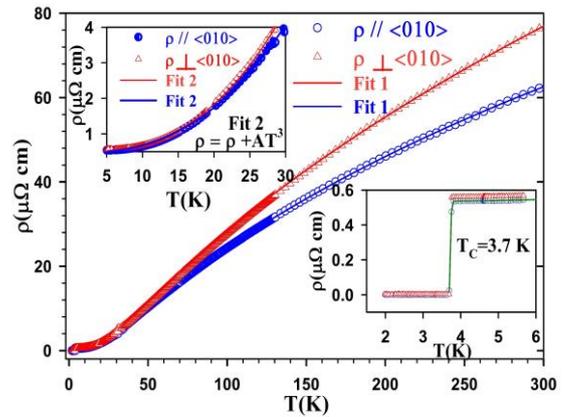

**Fig. 3.** Resistivity vs Temperature plot of BiPd for current parallel and orthogonal to <010> direction respectively.

High temperature part (100 K to 300 K) of normal state resistivity shows negative curvature with significant deviation from generally expected linear variation of resistivity with temperature. This can be explained and fitted (Fit 1) by parallel resistor model (already discussed in detail in Ref [7]) proposed by Wisemann et al.

## 3. Conclusion

We have probed the strength of anisotropy in BiPd via transport and magnetization measurements. Significant anisotropy has been found among some of the superconducting as well as normal state properties. Nearly 33 % anisotropy has been probed in upper critical field (Hc$_2$) at 1.8 K. Anisotropy gradually builds up below 200 K in case of normal state susceptibility which reduces at low temperatures (<5 K). Resistivity measurements has revealed gradual growth of anisotropy with increasing temperature in BiPd and eventually reaches to 30 % at room temperature. These anisotropy effects possibly arise due to the significant Rashba type antisymmetric spin orbit coupling present in NCS α-BiPd. These results are in agreement with our recent band structure calculations.

## Acknowledgment

We would like to acknowledge Mr. Anil Kumar and Mr. Ulhas Vaidya for their help in transport and magnetization measurements.

## References

[1] L. P. Gorkov and E. I. Rashba, Phys. Rev. Lett. **87**, 037004 (2001).
[2] E. Bauer, G. Hilscher, H. Michor, Ch. Paul, E. W. Scheidt, A. Gribanov, Yu. Seropegin, H. Noe¨el, M. Sigrist, and P. Rogl, Phys. Rev. Lett. **92**, 027003 (2004).
[3] H. Q. Yuan, D. F. Agterberg, N. Hayashi, P. Badica, D. Vandervelde, K. Togano, M. Sigrist, and M. B. Salamon, Phys. Rev. Lett. 97, 017006 (2006).
[4] Y. Tanaka, Y. Mizuno, T. Yokoyama, K. Yada, and M. Sato, Phys. Rev. Lett. **105** (2010) 097002.
[5] X.-L. Qi and S.-C. Zhang, Rev. Mod. Phys. 83 (2011) 1057.
[6] Shunji Matsuura, Po-Yao Chang, Andreas P. Schnyder, Shinsei Ryu, New Journal of Physics 15, 045019 (2013)
[7] B. Joshi, A. Thamizhavel, and S. Ramakrishnan, Phys. Rev. B **84** (2011) 064518.
[8] M. Mondal, B. Joshi, S. Kumar, A. Kamlapure, S. C. Ganguli, A. Thamizhavel, S. S. Mandal, S. Ramakrishnan, and P. Raychaudhuri, Phys. Rev. B **86** (2012) 094520.
[9] Kazuaki Matano, Satoki Maeda, Hiroki Sawaoka, Yuji Muro, Toshiro Takabatake, Bhanu Joshi, Srinivasan Ramakrishnan, Kenji Kawashima, Jun Akimitsu, Guo-qing Zheng, J. Phys. Soc. Jpn. 82 084711 (2013)
[10] A. H. Wilson, Theory of Metals (Cambridge University Press, Cambridge, England, 1958)